\documentstyle[pra,aps]{revtex}
\begin{document}
\def\question#1{{{\marginpar{\tiny \sc #1}}}}
\draft
\title{Cosmological Challenges in Theories with Extra Dimensions
and Remarks on the Horizon Problem}
\author{Daniel J. H. 
Chung\thanks{Electronic mail: {\tt
djchung@umich.edu}} and Katherine Freese\thanks{Electronic mail: {\tt
ktfreese@umich.edu}}} 
\address{Randall Physics Laboratory,
        University of Michigan, Ann Arbor, MI 48109-1120}
\date{April 1999}
\maketitle
\begin{abstract}
We consider the cosmology that results if our observable
universe is a 3-brane in a higher dimensional universe. In
particular, we focus on the case where
our 3-brane is located at the $Z_2$ symmetry
fixed plane of a $Z_2$ symmetric five-dimensional spacetime, as in the
Ho\v{r}ava-Witten model compactified on a Calabi-Yau manifold.  As
our first result, we find that there can be 
substantial modifications to the standard 
Friedmann-Robertson-Walker (FRW) cosmology; as a consequence,
a large class of such models is observationally inconsistent.  
In particular, any relationship between the Hubble constant
and the energy density on our brane is possible, including (but not only)
FRW. Generically, due to the existence of
the bulk and the boundary conditions on the orbifold fixed plane,
the relationship is not FRW, and hence
cosmological constraints coming from big bang nucleosynthesis,
structure formation, and the age of the universe difficult to satisfy.
We do wish to point out, however, that some specific choices
for the bulk stress-energy tensor components do reproduce
normal FRW cosmology on our brane, and we have constructed
an explicit example.  As our second result,
for a broad class of models, we find a somewhat surprising fact:
the stabilization of the radius of the extra dimension and hence the
four dimensional Planck mass requires unrealistic fine-tuning of the
equation of state on our 3-brane.  In the last third of the paper, we
make remarks about causality and the horizon problem that apply to
{\it any} theory in which the volume of the extra dimension determines
the four-dimensional gravitational coupling.  We point out that some
of the assumptions that lead to the usual inflationary requirements
are modified.
\end{abstract}
\def\question#1{{{\marginpar{\tiny \sc #1}}}}
\def\eqr#1{{Eq.\ (\ref{#1})}}
\def\be{\begin{equation}}
\def\ee{\end{equation}}

\section{Introduction}
Recently, there has been a great deal of renewed interest in the physics of
extra dimensions, the existence of which is a generic feature of
string theories.  In this paper,
we consider the cosmology that results if our observable
universe is a 3-brane in a higher dimensional universe. In
particular, we focus on the case where
our 3-brane is located at the $Z_2$ symmetry
fixed plane of a $Z_2$ symmetric five-dimensional spacetime;
our work applies to any 3-brane with this $Z_2$ symmetry.
Hence this work applies to a
class of models proposed by Ho\v{r}ava and Witten \cite{horavawitten}
which compactify M theory on an orbifold $S^1/Z_2$, with one of the
$E_8$ on one orbifold fixed plane and the other $E_8$ on the other
orbifold fixed plane.  The size of one of the extra dimensions
is determined by the orbifold radius (we will call this
extra dimension `the bulk').
In the Ho\v{r}ava and Witten models, once
the rest of the extra dimensions are compactified
on a Calabi-Yau manifold with a radius smaller than the orbifold
radius, the standard model (SM) fields charged under a subgroup of
$E_8$ are confined to one of the two 3-branes while gravity is free to
propagate in the bulk.  There has been some effort
to explore the cosmologies resulting
from such a class of models which we will refer to as the
Ho\v{r}ava-Witten models.  
Some of this work has focused on the inflationary periods or on the
modifications to the particle physics related phenomena as a result of
Kaluza-Klein excitations.  There has been additional study
of cosmology in scenarios without a $Z_2$ symmetry.  References
of previous cosmological work in the context of our observable
universe as a 3-brane in a higher dimensional context include
\cite{cosmobefore,chamblin,ovrut}.

In this work we focus on rather general cosmological considerations
of a 3-brane with $Z_2$ symmetry.  We explore consequences of
the main novel features: i) the
boundary conditions on the orbifold fixed planes induced by the $Z_2$
symmetry and ii) the dependence of the
4-dimensional Planck constant on the
orbifold proper length.  In the first two-thirds of this paper, we
restrict ourselves to models in which the spacetime has the structure
${\cal M}_4 \times S^1/Z_2$ where ${\cal M}_4$ is foliated by flat,
homogeneous and isotropic 3-slices and find that, generically,
substantial modifications to the observable Friedmann-Robertson-Walker
(FRW) universe result.  In the last third of the paper, we make
remarks about causality and the horizon problem that apply to {\it
any} theory in which the volume of the extra dimension determines the
four-dimensional gravitational coupling.  While this work was in
progress, some of the same results have been presented by
\cite{binetruy}; however, much of our work covers additional aspects
of cosmology in theories with extra dimensions as well\footnote{Also,
as this work was being submitted, we became aware of \cite{csaki}
and \cite{cline2}, which also discuss brane cosmology.}.

Because the novel elements of Ho\v{r}ava-Witten models result in a
different set of cosmological equations, observational constraints can
be imposed on such models.  We find that, from a five dimensional
point of view in general, FRW cosmologies are not the generic
solutions on the brane containing our world.  In particular, there is
no manifest dynamical mechanism which drives the the theory to have
the fields living on the brane dominating the stress energy.  Hence,
one rarely recovers the standard relationship $H^2 \sim \rho/m_{pl}^2$
where $\rho$ is the energy density of the fields living on the brane
and $H$ is the Hubble speed on the 3-brane.  The main point is that
the behavior of $H$ and thus the evolution of our observable universe
is controlled by the moduli\footnote{In this paper, we use the word
moduli to refer to any light field degrees of freedom that contributes
to the effective stress energy tensor.}  from the extra dimensions and
the attending boundary conditions of the spacetime, rather than by
what is on our brane. This statement is true even when the bulk is
``empty.'' One can think of this as a version of the moduli problem,
which states that our 3-brane's stress energy can be generically
subdominant compared to the effect of the moduli (particularly those
not confined to our 3-brane) on our 3-brane's cosmological history.
It is possible although not generic
to reproduce ordinary FRW on our brane, and we construct a 
specific example in which FRW does result.
Although we discuss this problem in the context of a compact $S^1/Z_2$
extra dimension motivated by the Ho\v{r}ava-Witten model, this
conclusion should apply to any theory in which our observable universe
is a 3-brane located at the $Z_2$ symmetry fixed point of a $Z_2$
symmetric five dimensional spacetime.

The altered relationship between H and $\rho$ leads to drastic
modifications of the results from Big Bang Nucleosynthesis.  The
temperature/time relationship at the time of Big Bang Nucleosynthesis
is drastically modified, and the element abundances produced in a
generic model are in violation of observation.  
We also discuss the growth of density perturbations required
for large scale structure formation and find that the standard results
are no longer obtained: galaxy formation cannot proceed in the usual
fashion.

We also find the result that, in a broad class of models, it is
generically impossible to stabilize the radius of the orbifold without
causing the equation of state $w = P/\rho$ on our brane to diverge;
{\it i.e.,} fine tuning of the equation of state is required in order
to allow one to fix the four dimensional Planck mass today at a
constant value.  Because the 3-brane observable universe that we live
in has an energy density whose scale is much smaller than the
fundamental Planck scale, we would naively expect that the stress
energy of the 3-brane is irrelevant as far as the stabilization of the
orbifold is concerned.  However, this need not be true in general
since the brane stress energy must be consistently coupled to the bulk
stress energy.  We give an example where an absolutely static orbifold
is physically forbidden due to 1) this bulk/boundary coupling and 2)
the fact that the equation of state on the boundary is constrained to
satisfy $-1< w = P/\rho < 1$.  In this class of models, we find that
without fine tuning of the equation of state to $w=-2/3$ or without
having just a cosmological constant type of equation of state $w=-1$,
it is impossible to ``pin'' the four dimensional Planck mass to
today's value; i.e., it is impossible to stabilize the radius of the
orbifold.  Instead, the four dimensional Planck mass must still be
changing, albeit slowly, in order to satisfy the constraints and the
equations of motion.
We find that even when the Newton's constant is
allowed to vary slowly (within experimental bounds), it is difficult
to construct a realistic model that does not contain an unnaturally large
number.  The large number arises from the mismatch between the energy
density on our 3-brane and the energy density set by the scale of the
fundamental Planck's constant, i.e., a version of the cosmological
constant problem. Although we have only shown this difficulty with
stabilizing the radius of the extra dimension (and hence the
four-dimensional Planck mass) explicitly for specific models, we
suspect that this difficulty may persist more generally for other
models with extra dimensions.

We also make some remarks about causality and the solution to the
horizon problem in theories with extra dimensions.  These remarks
apply to any theory, regardless of boundary conditions, in which the
observable universe lies on a three-brane and gravity is described by
an extra dimension.  We caution that many assumptions go into the
standard inflationary requirements, and illustrate how a few of these
assumptions may be modified here.  In particular, the assumption that
$H^2 \sim T_c^2 /m_{pl}(t_c)$ at an early time may be modified. In
addition, the fact that the value of the four dimensional Planck mass
changes as the
radius of compactification changes must not be forgotten in inflation
models, and in fact may be exploited to solve the horizon problem
(see also \cite{riotto}).

In Section II, we present the model with a discussion of boundary
conditions.  In Section III, we discuss the
resulting nonstandard cosmology on our brane.  In Section IV, we
explore constraints from nucleosynthesis, structure formation, and the
age of the universe in the context of the nonstandard cosmology. In
Section V we show that, in a broad class of models, the four
dimensional Planck mass cannot be fixed at a constant value. In
Section VI we discuss causality and the horizon problem. We conclude in
Section VII.

\section{Model}

In the first part of this paper we restrict ourselves to a scenario in
which standard model particles are confined to one of the two orbifold
fixed planes (3-branes) and gravity resides in the bulk.  We consider
a space-time with the structure $S^1/Z_2 \times {\cal M}_4$ where
$S^1/Z_2$ is an orbifold and ${\cal M}_4 $ is a smooth manifold.  
We use the metric convention $+, -, -, -, -$.  Let the action be
decomposed as
\begin{equation}
\label{eq:action}
S_5 = \frac{-1}{2 \kappa_5^2} \int d^5 x  \sqrt{g} {\cal R} + S_{orb}
+ S_{boundary} + 
S_{GH}
\label{eq:totaction}
\end{equation}
where ${\cal R}$ is the Ricci scalar in 5 dimensions, 
$g$ is the absolute value of the
determinant of the five dimensional metric $g_{MN}$, $\kappa_5$ is the
five-dimensional Newton's constant
($\kappa_5^2 \equiv 1/m_{pl,5}^3$), $S_{orb}$
represents the orbifold (or the bulk) action, $S_{boundary}$ represents the
boundary (orbifold fixed plane) action (i.e. integrated only over the
boundary), and  
$S_{GH}$ is the Gibbons-Hawking boundary term to be discussed below.

\subsection{Metric and Einstein Equations}

We consider the metric of the form
\begin{equation}
\label{eq:metric}
ds^2= e^{2 \nu(\tau, u)} d\tau^2 - e^{2 \alpha(\tau, u)} d{\bf x}^2 -
e^{2 \beta(\tau, u)} du^2
\end{equation}
which foliates the space into flat, homogeneous and isotropic spatial
3-planes.  Here ${\bf x} = x_1,x_2, x_3$ are the coordinates on the
spatial 3-planes while $u$ is the orbifold coordinate (this metric was
considered for example by Lukas, Ovrut, and Waldram \cite{ovrut}
within a similar context).  Let $S^1$ be parameterized by $u \in[-
l,  l]$ with the endpoints identified.  The $Z_2$ symmetry takes $u
\rightarrow -u$ and leaves two three dimensional planes fixed.  These
orbifold fixed planes then are at $u=0$ and at $u= l$.  In this
paper we will refer to these as the 3-branes. As in the
Ho\v{r}ava-Witten model \cite{horavawitten}, one of these 3-branes is
our observable universe (which we choose without any loss of
generality to be at $u=0$) and the other is a hidden sector that
communicates only gravitationally with our visible sector.  Thus we
define $a(t(\tau))=e^{\alpha(\tau,u=0)}$ as the scale factor
describing the expansion of the 3-brane that is our observable
universe where $t(\tau) \equiv \int d\tau \exp(\nu(\tau,u=0))$ is the
proper time of a comoving observer (with the brane's position 
fixed at u=0; but see below).  The four dimensional Newton's constant is
inversely proportional to the distance between these 3-branes,
\begin{equation}
\label{eq:one}
G_N \sim {1 \over 16 \pi m_{pl,5}^3 R(\tau)} \, 
\end{equation}
where 
\begin{equation}
R(\tau) \equiv \int_0^l \exp( \beta(\tau, u)) du \, .
\label{eq:orblength}
\end{equation}
Here $e^{ \beta (\tau, u)}$ is the scale factor describing the
expansion of the orbifold.

Note that we can always reparameterize the theory such that the
3-brane positions are fixed in orbifold coordinate space.  However, in
that case, we can no longer have the full freedom of gauge in the
$(\tau, u)$ plane.  For example, we can no longer make a coordinate
transformation such that $\nu(\tau, u)= \beta(\tau, u)$
(conformally flat in the $\tau$-u subspace), because such
a transformation generally will give the 3-branes a time dependence
for their position along the orbifold.  However, since we can always
reparameterize $u$ by itself or $\tau$ by itself, we can rewrite any
metric having the property $\beta - \nu = j(u)$ or $\beta - \nu =
v(\tau)$ (where $j(u)$ and $v(\tau)$ are real functions) in a conformally
flat form (in the $\tau-u$ subspace) even with 3-brane coordinate
positions fixed.

By varying the action \eqr{eq:totaction} we find that the bulk
equations are
\begin{eqnarray}
e^{-2 \nu} T_{00} & = & 3 e^{-2 \beta} ( \alpha' \beta' - \alpha'' -2
\alpha'^2) + 3 e^{-2 \nu} (\dot{\alpha}^2 + \dot{\alpha}\dot{\beta} )
\label{eq:t00} \\
e^{-2 \alpha} T_{11} & = & e^{-2 \beta} (3 \alpha'^2 -2 \alpha' \beta'
+ 2 \alpha' \nu' - \beta' \nu' + \nu'^2 + 2 \alpha'' + \nu'') +
\nonumber \\
& &  e^{-2 \nu} ( -3 \dot{\alpha}^2 -2 \dot{\alpha}\dot{\beta} -
\dot{\beta}^2 + 2 \dot{\alpha}\dot{\nu} + \dot{\beta}\dot{\nu} - 2
\ddot{\alpha} -  \ddot{\beta})  \nonumber \\
e^{-2 \beta} T_{44} & = &-3 (\alpha'^2 + \alpha' \nu') e^{-2 \beta} +
 3 e^{-2 \nu} ( 2  \dot{\alpha}^2 -\dot{\alpha}\dot{\nu} +
 \ddot{\alpha} ) \nonumber \\
T_{4 0} & = & 3 \nu' \dot{\alpha} + 3 \alpha' ( - \dot{\alpha} +
 \dot{\beta}) - 3 \dot{\alpha'}
\end{eqnarray}
where the dots denote partial derivatives with respect to $\tau$ while the
primes denote partial derivatives with respect to $u$.  

An alternative and equivalent way to think about the five-dimensional
space is a boundary picture.  Here the coordinate $u$ is restricted to
one half of the circle, $u \in [0, l]$ and five dimensional spacetime
is written as $[0, l] \times {\cal M}_4$.  The orbifold fixed
planes then turn into boundaries of this five-dimensional space.

We will take the 3-branes to be at fixed orbifold coordinate
$u$.  However, the general results of our paper, including the result that
ordinary FRW cosmology is modified by the existence of the bulk,
would qualitatively
hold even if one were to allow the branes to ``bend", i.e.,
give the location of the 3-branes some 
$u$-dependence.

\subsection{Boundary Conditions}

The bulk equations only apply in an open region that does not include
the boundary.  By continuity, the metric on the boundary (brane) is
determined by the metric in the bulk.  
It is the Israel conditions which connect the boundary and the bulk.
Hence, the bulk stress energy cannot be set to zero without imposing
strong constraints on the energy density of the boundary.

{\it General remarks on Israel Conditions:} Here we briefly remind the
reader of the Israel conditions (see \cite{chamblin} and references
therein), which are relevant when there are boundaries (or
equivalently discontinuities of the derivatives of the metric).
Alternatively, one can derive equivalent conditions by just
inserting a $\delta$ function for the matter action (see below) as is
done in \cite{ovrut,binetruy}.

The Gibbons-Hawking term in \eqr{eq:action} above takes the form
\begin{equation}
S_{GH} = \frac{1}{\kappa_5^2} \sum_{\mbox{faces}} \int_{\mbox{face}} d^4 x \sqrt{h} K
\end{equation}
where $K \equiv h^{M N} K_{MN}$ is the trace of the extrinsic
curvature $K_{M N} \equiv h^P_M \nabla_P n_N$, $h_{M N} \equiv g_{M N}
+ n_M n_N$ is the induced metric on the boundary surface, and
the vector $n^N$,
orthonormal to the surface, points toward the region which the surface
bounds.\footnote{The index $M$, for example, takes integer values 
from 0 to 4.  For example, if we work in the coordinate frame
of eqn. (2) with the boundary at a fixed coordinate value, then
the fact that $n_m n_N g^{MN} = 1$ implies that $n_4 = \sqrt{-g_{44}}
= e^{-\beta}$. Thus $h_{44} = g_{44} + n_4^2 = 0$, while $h_{MN}$
can be nonzero for $M,N < 4$, so that $h_{MN}$ indeed projects
onto the boundary.  It is then easy to see that only the tangential derivatives
(P = 1,2,3) survive in obtaining $K_{MN}$.}

Upon
varying $S_5$ with respect to the metric we find, in 
addition to the usual
\begin{equation}
{\cal R}_{M N} - 1/2 g_{M N} {\cal R} = T_{MN}
\end{equation}
(where $T_{MN} = \frac{2 \kappa_5^2}{\sqrt{g}} \frac{\delta
S_{orb}}{\delta g^{M N}}$)
within the bulk (which was used in the last subsection), the boundary variation
\begin{equation}
\int \frac{\delta S_{boundary}}{\delta g^{MN}} \delta g^{MN} + \int
\frac{\delta 
S_{GH}}{\delta g^{MN}} \delta g^{MN} + \frac{-1}{2 \kappa_5^2} \int
d^5 x \sqrt{g} \delta {\cal R}_{MN} g^{MN}=0
\end{equation}
where the last integrand is a total divergence and becomes a surface term.
This leads to the Israel conditions
\begin{equation}
\sum_{\pm \mbox{faces}} \left( K_{MN} - K h_{MN} \right) = t_{MN} \, ,
\label{eq:israel}
\end{equation}
where the sum over faces is for each side of the boundary surface and
we have defined
\begin{equation}
t_{MN}= \frac{2 \kappa_5^2}{\sqrt{h}} \frac{\delta S_{boundary}}{\delta h^{MN}}
\end{equation}
as the energy momentum tensor on the boundary.  We will assume that
this energy momentum tensor on the boundary can be written in a
perfect fluid form.  For example, 
\begin{equation}
\label{eq:energydensity}
t^0_0 = \kappa_5^2 \rho
\end{equation}
 and 
\begin{equation}
\label{eq:pressure}
t^1_1= - \kappa_5^2 P \, ,
\end{equation}
where $\rho$ and $P$ are the energy density and
pressure, respectively, measured by a comoving observer.
One could write the 00
component of eqn. (10) in the following way, if one also uses
eqn. (12). Defining $H_{in} = {1 \over a}{da \over dv}$ where
$dv = e^\beta du$ (by analogy with the Hubble constant),
we can write the 00 Israel condition as
$\rho = {3 \over \kappa_5^2} \sum_{\pm \mbox{faces}} H_{in}$.

The Israel condition in eqn. (10) is very similar
to $\Delta E_{perp} = 4 \pi \sigma$ in electrostatics,
where $\sigma$ is the charge/area on a charged plate and
$E_{perp}$ is the component of the electric field perpendicular
to the plate.  Eqn. (10) intimately relates the energy-momentum
on our brane to its extrinsic curvature, i.e., to the way
that it is imbedded in the bulk, just as the charge
on a plate is intimately connected to the electric
field on either side.  

Note that there is an alternative way of looking at the problem
which gives the same results.
Instead of partitioning the orbifold into regions with boundaries
(an approach adopted from \cite{chamblin}), one could elevate the integrand
of the surface action to a density in five dimensions by inserting a
$\delta$ function.  In that case, we would obtain
\begin{equation}
T_{MN}^{\mbox{total}}= T_{MN} + t_{MN} \delta(\Sigma)
\end{equation}
where $\Sigma$ represents the boundary surface with the appropriate
measure for the $\delta$ function built in. 
In lieu of eqns. (8) and (10) we would then use
${\cal R}_{MN} - {1 \over 2} g_{MN} {\cal R} = T_{MN}^{\mbox{total}}$
everywhere.  

Note also
that the Israel conditions \eqr{eq:israel} probe the global structure
of the spacetime even though each of the conditions is local.  They
probe the global structure because they result from the boundary terms
arising from integration over all space.  Hence, loosely speaking, the
topological constraints of the $S^1/Z_2$ are encoded into local
boundary conditions by the Israel conditions.

Note that no additional information, beyond what is contained
in Einstein's equations and the Israel conditions, is obtained
from considering the Bianchi identities.
In the bulk, the Bianchi identities automatically give
$\nabla_M T^{MN}=0$.\footnote{If the bulk contains only one field (say
a scalar  
field), then its equation of motion will be determined by the Bianchi
identities; if there are more fields, then other equations of motion
are needed as well.}  On the
boundary, $\nabla_M t^{MN} = \sum_{\pm \mbox{faces}} h^{NM} T_{MP}
n^P$ allows the energy to flow between the boundary and the bulk.
Since these equations follow from the Einstein equations and from
\eqr{eq:israel}, there are no additional constraints coming from these
equations.

\bigskip
{\it Israel conditions for our model:}
We can now find the Israel conditions specific to the metric
in \eqr{eq:metric}.  Our boundary conditions are found by considering
the two 3-branes at $u= 0$ and $u= l$.
These are the planes at which we
have discontinuities in the normal derivatives of the metric, and
hence where the Israel conditions for the boundaries give nontrivial
constraints.  

As mentioned above, we can always reparameterize the theory
such that the 3-branes are fixed in coordinate space. 
Because we are considering
a brane at a $Z_2$ symmetry fixed plane, from the sum
over faces in eqn. (10) we get two identical terms,
merely resulting in a factor of two.  In a more general spacetime,
the Israel conditions in eqn. (10) would still hold, but
one would have to explicitly add the contributions from
the two sides of the boundary surfaces since these contributions
would no longer be equal.  
With the $Z_2$ symmetry, the Israel conditions are
\begin{eqnarray}
\mp 6 \alpha'|_{\pm} & = &  e^{\beta - 2 \nu} t_{00} 
\label{eq:israelalpha} \\
\pm \nu'|_{\pm} & = & \frac{1}{2} e^{\beta - 2 \alpha} t_{ii} +
\frac{1}{3} e^{\beta - 2 \nu} t_{00}
\label{eq:israelnu}
\end{eqnarray}
where the $|$ implies that the equation is to be evaluated
at one of the two sides of the boundary surface.  In each equation,
the upper signs are to be used for the face at the boundary plane at
$u=0^+$ while the lower signs correspond to the face at the boundary
plane at $u= l^-$.\footnote{Here, for example, $u=0^+$ corresponds
to approaching $u=0$ from the positive side ($u>0$).}  Note that
because of
the $Z_2$ symmetry defining the orbifold fixed planes, the lower sign
also applies to the the surface $u=0^-$, for example.  We have also
denoted the boundary parallel spatial coordinates $\bf x$ with the
indices $i$. Using eqn. (12) and $g_{00} = e^{2\nu}$, 
we see that eqn. (15) can be written
$\rho = 6 \alpha' m_{pl,5}^3 e^{-\beta}$.

As mentioned above, if one makes a particular gauge choice such as
$\nu(\tau,u) = \beta(\tau,u)$, then one must allow the possibility of
a moving 3-brane.  Only if $\beta - \nu$ is independent of time or
independent of $u$ can one choose $\nu(\tau, u) =\beta(\tau,u)$ with
fixed brane positions without any loss of generality.  The Israel
conditions for a 3-brane having a time dependent coordinate $r(\tau)$
are
\begin{eqnarray}
 t_{00} & = & \mp 6 e^{2 (\nu -\beta)}  \left[\frac{ (e^{2 \nu}
 \alpha' + e^{2
\beta} \dot{\alpha}\dot{r} )}{ ( e^{2 \nu} - e^{2 \beta} \dot{r}^2)
\sqrt{ e^{-2 \beta} - e^{-2 \nu} \dot{r}^2} } \right] \\
t_{11} & = & \pm e^{2 \alpha} \left[ \frac{ \ddot{r} + 2 e^{2 (\nu -
\beta)} \alpha' + \dot{r}^2 (\beta' - 2\alpha') + e^{2 (\nu-\beta)}
\nu' -2 \dot{r}^2 \nu' + 2 \dot{r} (\dot{\alpha} + \dot{\beta}) - 
e^{2(\beta-\nu)}\dot{r}^3 (\dot{\beta} + 2 \dot{\alpha})
-\dot{r} \dot{\nu} }{(e^{2 \nu} - e^{2 \beta} \dot{r}^2)
\sqrt{ e^{-2 \beta} - e^{-2 \nu} \dot{r}^2}}\right].
\end{eqnarray}
As explained before, with a proper choice of coordinates, the 3-brane
positions can be fixed ($\dot{r}=0$) at the coordinates that we have
chosen for all bulk coordinate time.  However, if we insist instead
that $\nu=\beta$, then this apparent loss of a functional degree
of freedom will reappear in the nonzero $\dot{r}$ except under certain
conditions previously discussed.

\section{ Nonstandard Evolution of our 3-brane}
>From the 4-dimensional point of view of a low energy observer living
on the 3-brane, the effect of the extra dimensions can be considered
as the coupling of some extra moduli fields to the boundary energy
density.  The distinguishing feature of the brane scenario is that
these extra fields arise from the gravitational coupling to the extra
dimensions and hence must obey the boundary equations
\eqr{eq:israelalpha} and \eqr{eq:israelnu}.  From \eqr{eq:t00}, not
only is the usual four-dimensional Friedmann equation relating the
energy density and the Hubble speed modified, but \eqr{eq:israelnu}
restricts the possible relationship even further.  In general, there
seem to be no restrictions of the stress energy tensor that drive the
energy density confined on the 3-brane to behave as $H^2$ (an explicit
illustrative example of this will be given later in the next section
V).  Hence, the relationship between the energy density of matter
confined to the brane and the expansion scale factor will generally be
different from the standard cosmology.

Let us consider
why one does not generically 
recover the standard evolution for our observable universe: i.e. why in
general $H^2$ will not be proportional to the energy density confined to
the observable sector 3-brane.
The main
point is that the behavior of $H$ and thus the time evolution of our
observable universe can be controlled by the moduli from the extra
dimensions and the attending boundary conditions of the spacetime,
rather than by what is on our brane.  One can think of this as
a version of the moduli problem, which states that our 3-brane's
stress energy is subdominant compared to the effect of the moduli
coming from extra dimensions
on our universe's history.

The above statement is most transparent when one considers the full
five dimensional Einstein's equations rather than the dimensionally
reduced 4-dimensional effective action.  For example, consider the
case where there is no stress-energy in the bulk, i.e. $T_{\mu \nu} =
0$.
Still, the behavior of the extra dimension is determined by the
3-brane stress energy $t_{\mu \nu}$, and the derivatives along the
extra dimensions of metric components act as energy sources for the
Friedmann equation governing the expansion of the observable 3-brane.
Consider, for example, \eqr{eq:t00}.  From a higher dimensional point
of view, the first of the equations in \eqr{eq:t00} clearly shows the contributions to $H$ that
are independent of $T_{00}$, the bulk energy density.  
For example, we can take $\beta$ to be a constant ($\dot{\beta} =
\beta'=0$) and set $T_{00}=\alpha''=0$.  
Then, using the Israel condition in
\eqr{eq:israelalpha}, as well as \eqr{eq:one}, 
\eqr{eq:energydensity}, and the fact
that $H = {da/dt \over a} = e^{-\nu} \dot{\alpha}$, we obtain the result
\begin{equation}
H= \frac{8 \pi G_N}{3} (\sqrt{2}  l \rho)  \, ,
\end{equation}
which shows a linear relationship between $\rho$ and $H$.  Hence, even
if the bulk is empty, we generally will not recover the standard FRW
cosmology.  This unusual relationship has also been discussed by Ref.\
\cite{binetruy}, which considered the 4-4 component of the Einstein's
equation with the assumption 
\begin{equation}
\rho_{bulk} \ll \frac{\rho_{brane}^2}{m_{pl,5}^3},
\end{equation}
where $\rho_{brane}$ is the energy density on the brane and
$\rho_{bulk}$ is the energy density on the bulk, and showed that an
unusual form of the Friedmann equation ensues (a linear
relationship between $H$ and $\rho$).

In general, almost any relationship between $\rho$ and $H$ is possible
(e.g. such as $\rho \sim H^q$).  One can see this, for example, by
using the fact that $\alpha''$ and $\beta'$ on the 3-brane are not
well constrained.  From the four dimensional effective theory point of
view, this is the same as saying that the moduli not confined to our
brane can contribute to the energy density such that nearly arbitrary
time dependence for the expansion scale factor $a$ may be achieved.
More explicitly, suppose we seek a solution of the form
\begin{equation}
\rho = \mu H^q
\end{equation}
where $\mu$ is a constant which we can parameterize as $M^{4-q} c$
where $M$ is a mass scale and $c$ is a dimensionless
constant.  In that case, the Israel equations, at $u=0$ for
example, become
\begin{eqnarray}
- \nu'(\tau, u=0) & = & 3 \alpha'(\tau, u=0) (w(\tau)+\frac{2}{3}) \\
 6 \alpha'(\tau,u=0) e^{-\beta(\tau,u=0)} &=& \mu \kappa_5^2
(e^{-\nu(\tau, u=0)} \dot{\alpha}(\tau,u=0))^q 
\end{eqnarray}
where $w=-t^i_i/t^0_0$ is the equation of state.  Note that because
these constraint equations are only evaluated at a point on the
orbifold, one can consider $\nu(\tau,u=0)$ and $\nu'(\tau,
u=0)$ as independent functions.  Similarly, $\alpha(\tau,u=0)$ and
$\alpha'(\tau, u=0)$ can be considered independent functions.  

For example, we can construct
a solution for any value of $q$ with $w\equiv P/\rho=0$ (pressureless
fluid) and $\rho \propto 1/a^3(\tau,u=0) = 
\exp(-3 \alpha(\tau, u=0))$ as follows:
\begin{eqnarray}
\beta(\tau, u) & = &\nu(\tau, u) = \frac{\kappa_5^2}{3} \left[ \frac{q}{3
\tau} \right]^q \mu u \label{eq:examplesol1} \\
\alpha(\tau, u) & =& \frac{-\kappa_5^2 \mu}{6}  \left[ \frac{q}{3
\tau} \right]^q ( F(u) - F(u=0)) + \mbox{ln}(\lambda) + \frac{q}{3}
\mbox{ln}(\frac{\tau}{\tau_0})
\label{eq:examplesol2}
\end{eqnarray}
where $F(u)$ is any smooth function satisfying $F'(u=0)=1$ and
$\lambda$, $\tau_0$, and $\mu$ are constants.  For example, suppose
$q=2$ (corresponding to the usual FRW relationship) and $F(u)=u$.  The
resulting stress energy tensor components are
\begin{eqnarray}
T^0_0 &= & \frac{- 4 \exp( \frac{-8 \kappa_5^2 \mu u}{27 \tau^2}) ( - 81
\tau^4 + 4 \kappa_5^4 \mu^2 (\tau^2 + u^2))}{243 \tau^6} \\
T^1_1 &=&  \frac{-4 \exp( \frac{-8 \kappa_5^2 \mu u}{27 \tau^2})
\kappa_5^2 \mu ( -36 \tau^2 u + \kappa_5^2 \mu (\tau^2 - 4 u^2))}{243
\tau^6} \\
T^4_4 &=&  \frac{-2 \exp(\frac{-8 \kappa_5^2 \mu u}{27 \tau^2}) (81
\tau^4 + 54 \kappa_5^2 \mu \tau^2 u + 2 \kappa_5^4 \mu^2 (\tau^2 + 16
u^2))}{243 \tau^6} \\
T_{4 0} & = & \frac{40 \kappa_5^4 \mu^2 u}{243 \tau^5} \\
\rho & = & \frac{4 \mu}{9 \tau^2} \\
P & = & 0
\end{eqnarray}
Hence, with a nontrivial time dependent bulk stress energy tensor in
the bulk, one can recover the FRW expansion on the boundary.  Note
that in this particular example, there is no energy flowing between
the brane and the bulk.

Of course, once a specific model is written down with specific choices for
the stress-energy tensor components $T_{\mu\nu}$, we will be able
to solve Einstein's equations together with the boundary conditions
to calculate exactly which $q$ results in $\rho \sim H^q$.  At present
we can only point out that different choices of $T_{\mu\nu}$ will
give rise to different values of $q$, and generically will produce
modifications to the standard FRW results.

Let us now consider the modifications to the standard
FRW cosmology from a purely four dimensional effective action point of
view.  The total action, \eqr{eq:totaction}, can be
dimensionally 
reduced to
\begin{equation}
S_4 \approx \frac{-R}{2 \kappa_5^2} \int d^4 x  \sqrt{g_4} 
({\cal R}_4 + \Delta {\cal R}) + S_{orb 4} + 
 S_{boundary 1} + S_{boundary 2}
\label{eq:dimred}
\end{equation}
where $\Delta {\cal R}$ is the gravitational moduli contribution that
arises from dimensionally reducing the 5-dimensional Ricci scalar,
$S_{orb 4}$ is the contribution coming from the bulk field, and the
remaining action terms correspond to the brane-confined field
contributions.  In the Ho\v{r}ava-Witten model, $S_{boundary 1}$
corresponds to the contribution from the observable sector and
$S_{boundary 2}$ corresponds to the contribution from the hidden
sector.  What we are calling the moduli problem is that $S_{boundary
2}$, $S_{orb 4}$, and $\Delta {\cal R}$ contributions can dominate the
energy density determining the expansion rate of the universe.  For a
particular orbifold geometry about which the dimensional reduction is
taken, the stress energy contributions coming from $\Delta{\cal R}$,
$S_{boundary 1}$, $S_{boundary 2}$ and $S_{orb 4}$ are related through
the Israel conditions and the Einstein equations.  In Ref.\
\cite{binetruy}, a linear relationship between $H$ and $\rho$ arising
from this moduli dominance was emphasized, particularly when $S_{orb
4}=0$.  We would like to merely point out that in general, any
relationship between $H$ and $\rho$ may result if all the moduli
contributions (i.e. $\Delta{\cal R}$, $S_{boundary 2}$, and $S_{orb
4}$) are taken into account.

\section{Constraints}
In the last section, we saw that the ordinary matter confined
to our brane does not (alone) determine the expansion rate of our
universe $H$; in contrast the dilution of the energy density and the
pressure of the ordinary matter proceeds as usual as the universe
expands.  Hence, any physical observable that measures the expansion
rate independently of the ``ordinary'' energy density will be able to
test this 3-brane scenario.  To illustrate this constraint, we will
consider big bang nucleosynthesis (BBN), structure formation, and the
age of the universe.

{\it Big Bang Nucleosynthesis:} As discussed above, generically one
obtains modifications to the cosmological standard model relationship
between evolution of the scale factor and energy density on our brane.
These modifications will drastically alter nucleosynthesis.  In BBN, a
given reaction rate depends directly on the energy density (the photon
number density, for example) while the freeze out is governed by
the ratio of  the Hubble speed to the reaction rate. Hence, the energy
density time variation can be compared with the universe expansion
rate since the ratios of the various light element abundances measure
the various temperatures at which different elements freeze out.

At the high temperatures in the early universe, the ratio of neutrons
to protons is determined by its thermal equilibrium value,
\begin{equation}
n/p = e^{-Q/T} \,\,\,\,\,\, , T \geq T_F \, ,
\end{equation}
where the neutron-proton mass difference $Q=1.293$ MeV.  
Neutrons drop out of equilibrium below a freeze-out temperature 
$T_F$, where the weak interaction rates can no longer keep up
with the expansion of the universe.  Below $T_F$
the $n/p$ ratio continues to fall due to $\beta$-decay on the time
scale of the neutron half-life $\tau_n$.  In the standard BBN scenario,
nucleosynthesis begins at a temperature approximately
given by
\begin{equation}
\label{eq:deuterium}
T_D = 2.2 {{\rm MeV} \over - {\rm ln} \eta} 
\end{equation}
where $\eta$ is the baryon to photon ratio.
Once $T_D$ is reached, deuterium becomes stable against
photodissociation and nucleosynthesis takes place very
rapidly, efficiently converting essentially all of
the available neutrons into He$^4$.  In this approximation,
the primordial helium abundance $Y_p$ is given by
\begin{equation}
\label{eq:yp}
Y_p = \bigl( {2n \over n+p} \bigr)_D =
\bigl( {2n \over n+p} \bigr)_F {\rm exp}[-\Gamma(t_D-t_F)]
\sim { 2 e^{- \Gamma t_D} \over 1+ {\rm exp[Q/T_F]}} \, 
\end{equation}
where the final approximation is valid since $\Gamma^{-1}
= \tau_n/{\rm ln} 2 \gg t_F \sim 1$sec.

Now we consider the modification of these results when the
equation $H^2 \sim \rho /m_{pl}^2$ is modified.
As an example, let us consider the following modification to standard
4-dimensional Einstein equations:
\begin{equation}
\label{eq:modi}
H = \rho / M^3 \, ,
\end{equation}
where $M$ is some mass scale, perhaps generically the Planck mass.
Such a relationship was discussed above and is also found in the toy
model discussed in the next section. It was also found by Lukas,
Ovrut, and Waldram \cite{ovrut} in their nonlinear case as well as by
\cite{binetruy} in a similar context.  If we assume that the scale
factor expansion is a power law, we then have $H \sim 1/t$.  In a
radiation dominated universe, $\rho \sim T^4$.  With \eqr{eq:modi},
one can then find that the scale factor grows as $a \sim t^{1/4}$ and
the modified temperature-time relation is $T \sim t^{-1/4} M^{3/4}$.
Recall that the freeze-out of neutrons and protons occurs when a
typical $n \leftrightarrow p$ weak interaction rate $\Gamma \sim G_F^2
T_F^5$ is equal to the expansion rate $H$.  Here $H \sim T^4/M^3$ (as
opposed to the usual $T^2/m_{pl}$ of the standard BBN model). Thus we
find
\begin{equation}
T_F = \bar{T}_F^3 (m_{pl}/M^3) \, ,
\end{equation}
where an overbar indicates the standard BBN model value.
For example, if $M=m_{pl} = 10^{19}$GeV, we have
$T_F = 10^{-44} \bar{T}_F$.  This model has the 
novel feature that $T_D > T_F$, so that the exponential
factor in the numerator of \eqr{eq:yp} is not there.
However, both temperatures are significantly lower than the usual $\bar{T}_F$, 
so that the production of He$^4$ is driven 
to zero!  

In addition to the specific case considered above of $H \propto \rho$,
one can allow almost any $q$ in the equation $\rho \propto H^q$,
depending on the $T_{\mu \nu}$ in the bulk.  To analyze a variety of
possibilities, we consider power law growth of the scale factor both
in the matter and the radiation dominated regimes, and allow the power law
to be a free parameter.  Thus we take the scale factor on the 3-brane
to be $e^{\alpha(\tau, u=0)} \sim (t(\tau, u=0)/t_i)^n$ for the radiation
dominated era, where $t_i$ is a constant and $t(\tau, u=0)$ is the
proper time of the comoving observer on the 3-brane. During matter
domination we take the power law index to be $m$ instead of $n$ (with
$m$ and $n$ usually different).  Consider the neutron-proton
interconversion interactions freezing out.  By setting the reaction
rate equal to the Hubble expansion rate, to within an order of
magnitude, the freeze out temperature then is found to be
\begin{equation}
T_{F} \sim \left( G_F^2 \frac{T_0^{1/m}}{H_0} (\rho_{0m} / \rho_{0r})
^{[1/n-1/m]}
\right)^{\frac{1}{(1/m -5)}}
\end{equation}
where $T_0$ is the temperature of the cosmic background photons today
$H_0$ is the Hubble speed today, $\rho_{0m}$ is the matter
density today, and $\rho_{0r}$ is the radiation density today.  

Compare for example the two possible cases: 1. $\{n=1/2, m=2/3\}$
(usual FRW scenario) and 2. $\{n=1/4, m=1/3\}$ (a possible brane
scenario).  For case 1, we obtain a crude estimate $T_{F} \sim 1$ MeV
while for case 2, we obtain a crude estimate of $T_{F} \sim 10^{14}$
GeV.  Hence, even if this crude estimate were orders of magnitude
off,\footnote{The actual freeze out temperature will be lower, making
the proton-neutron relative abundance ratio (which is approximately
$\exp(-1 MeV/T_{F})$) discrepancy from the observed smaller.} we will
obtain a large measurable discrepancy for the resulting element
abundances since the neutron to proton ratio is constrained to a
precision of less than $1 \%$.

{\it Structure Formation:} 
Consider the Jeans analysis of a fluid
overdensity within a universe with $a \sim t^m$ during
the matter dominated epoch.  To obtain an
idea of the type of effects that we may find, instead of analyzing the
exact perturbation equations with metric perturbations
included,\footnote{This more careful treatment is in progress.} we
will merely modify the time dependence of the scale factor in the
usual Jeans analysis equation valid for standard FRW cosmology. We
assume the usual long wavelength limit and consider only one component
nonrelativistic matter fluid.  Normalizing the background energy
density and the Hubble speed to approximately what we measure today
(i.e. $\rho_0 \approx 3/(8 \pi) M_{pl}^2 H_0^2$), the fluid
overdensity equation can be written as
\begin{equation}
\delta''(x) + \frac{2 m}{x} \delta'(x) -  \frac{3 m^2}{2} x^{-3 m} \delta =0
\end{equation}
where $\delta \equiv (\rho-\bar{\rho})/\bar{\rho}$ is the fluid
overdensity and $x\equiv t/t_0$ with $t_0$ denoting the time today.
This equation can generally be solved in terms of Bessel functions for
constant $m$.  In the standard FRW cosmology with matter domination,
$m=2/3$, and there is one solution to $\delta$ which grows as the
scale factor $\delta \sim a \sim t^{2/3}$ and the other decreasing as
$1/x$. However, take for
example the case where $\rho \sim H$ which is equivalent to setting
$m=1/3$ if $\rho \propto a^{-3}$.  In this case, there will be two
growing solutions for $\delta$, one having the form $1+x/4+ . . .$
and
the other $x^{1/3}(1+ x/8 + . . .)$.  This agrees with the expectation
that in a universe that is expanding more slowly, the overdensity will
grow more quickly with the scale factor.  

Extra growth of the perturbations compared to the standard picture
might enable some new models to be considered, that are ruled
out in the standard model. For example, one might be able to better
tolerate a smooth component in the universe (which
ordinarily suppresses perturbation growth) and still get enough
growth of the fluctuations to get structure today.  For example,
particles that decay to radiation might be able to lead to a longer 
radiation dominated universe subsequent to the initial matter/radiation
equality. However, the standard picture of cold dark matter
might well be in trouble with altered growth of the scale factor.
With this altered growth, calculation of the integrated Sachs Wolfe effect
and normalization to the COBE data would undoubtedly lead to
anisotropies in the power spectrum which are less scale invariant
than the standard model, such that one would not be able to match
the data on all different scales.

The time dependence of the scale factor for a given type of ordinary
energy density $\rho$ can be tested by considering the location
of where the collapse and diffusion processed power spectrum starts to
flatten out.  This flattening occurs in the standard CDM scenarios
because a particular nonrelativistic matter fluctuation Fourier mode
grows only after the matter domination phase begins and the mode
wavelength enters the horizon.  The horizon length at the time of
matter radiation equality is approximately $t_{eq}$.  For wavelengths
longer than $t_{eq}$ (at that time), the modes are outside of the
horizon and will not grow until they enter the horizon.  Hence, today,
the mode's amplitude will be suppressed with respect to the mode that
grew starting at time $t_{eq}$.  Modes that had already
entered the horizon by the time $t_{eq}$ will grow together, and hence the
spectrum will be flat in that region.  

With a different time dependence for the scale
factor $a \sim t^m$, the location of this bend in the power spectrum will
be
shifted from $\lambda_1$ to $\lambda_2$
\begin{equation}
\lambda_2= \frac{3 m}{2} \left( \frac{\rho_{0R}}{\rho_{0M}}
\right)^{\frac{1}{m} - \frac{3}{2}} \lambda_1 \, .
\end{equation}
For the case $\rho \propto H$ (m=1/3), we have $\lambda_2= 10^{-7}
\lambda_1$ if we
assume that $\rho_{0 M}$ is order of the critical density ({\it{i.e.}}, 
$8.1 \times 10^{-47} \mbox{GeV}^4$).

{\it Age of the Universe}
With the time dependence of the scale factor altered, the age of the
universe may be modified.  For example, if we assume that
structures form during a period after the
matter-radiation equality with $a \sim t^m$ and that the scale
factor at matter-radiation equality $a_{\mbox{eq}}$ is much smaller
than today's scale factor $a_0$, i.e. $(a_0/a_{\mbox{eq}}) \gg 1$, then
the time period of structure formation is given by $t=m/H$ where $H$ is
the Hubble speed today.  Hence the lower bound on the age of the
oldest stars in a globular clusters may be used to constrain $m$.
Taking the value of \cite{chaboyer} of $t > 10.2$ Gyr, we find
\begin{equation}
m>h
\end{equation}
where the Hubble speed is $H=100 h$ Km/s/Mpc.  This would rule out
$\rho \propto H$, since in that case $m=1/3$, whereas
observations of the Hubble constant find that $h > 0.5$.  

\section{Difficulties of Stationary Planck Mass Cosmology}

We have discussed the altered relationship between the
energy density and the Hubble speed in the 3-brane scenario compared to 
the standard cosmology. Another difference from the standard
cosmology is that Newton's gravitational coupling constant can have a
time dependence in the Ho\v{r}ava-Witten scenario because it is 
inversely proportional to the orbifold proper length.  Naively,
one would, in
constructing a cosmological scenario, then try to freeze the orbifold
radius.  
However, as we will see, this turns out to be unphysical, at least in
the case where the metric can be written in a conformally flat form in
the $\tau-u$ subspace.
If one allows the orbifold length to vary slowly but within the
limits of astrophysical bounds, we still expect some constraint
although it is not as severe.  We will explicitly illustrate this with
a toy model.  
This difficulty of freezing the four dimensional Planck's constant is yet
another manifestation of the cosmological constant problem.

We will consider the stress-energy tensor in the comoving frame.  
As in \eqr{eq:energydensity} and \eqr{eq:pressure}, we 
denote the energy density and the pressure as $t^0_0 = \kappa_5^2
\rho$ and $t^i_i = - \kappa_5^2 P $ respectively.  Hence,
\eqr{eq:israelnu} can be rewritten as
\begin{equation}
\nu'|_\pm  =  \pm  (\frac{1}{2} w + \frac{1}{3}) t^0_0 e^{\beta}
\label{eq:nuconst}
\end{equation}
where we have introduced a time dependent function $w \equiv P/\rho$
defining the equation of state for the stress energy.  Now, suppose we
have a metric with
the property that $\nu - \beta$ is purely a function of $u$
or $\tau$; any such metric can be put into the conformally flat form in
the $\tau-u$ subspace. In other words, for any such metric
we can effectively take $\nu = \beta$ in Eq.(2).  Given such a metric,
we will now attempt
to fix the orbifold radius and consequently the four-dimensional
Planck mass by taking $\beta$ to be time independent. 
Since
\eqr{eq:nuconst} holds for all time, we can
impose the condition
\begin{equation}
\frac{d}{dt} \beta'|_\pm =0
\label{eq:staticbulk}
\end{equation} 
where, since $\beta$ is time independent, $t= e^{\beta(u)} \tau$.
Solving \eqr{eq:staticbulk} and \eqr{eq:nuconst} for $w(t)$, we find
\begin{equation}
w(t) = \frac{-2}{3} + \frac{w(t_i) + 2/3}{(\rho(t)/\rho(t_i))} \, ,
\label{eq:naive}
\end{equation}
where $t_i$ represents some initial time.  

Hence, for all energy
densities that decrease with time and $w(t_i) \neq -2/3$, our equation
of state grows in magnitude without bound.  For the usual macroscopic
stress energy consisting of homogeneous scalar fields and fermionic
particles bound to the brane, the equation of state is constrained to
$-1 < w(t) < 1$.  Hence, \eqr{eq:naive} most likely will violate any
realistic equation of state, and the four dimensional Planck mass
cannot be fixed to a constant value (e.g. with a potential). 
We comment briefly on the possibility that $w(t) = w(t_i) = - 2/3$
exactly, 
the only value that allows the radius to stabilize for decreasing
energy density.  We note that this value has been proposed 
(Wang {\it et al} \cite{wang}) as the best fit
value to the combination of a number of data sets, including recent
supernovae observations as well as various measures of large scale 
structure and the microwave background. However, in the context of
\eqr{eq:naive}, obtaining this value would require either some
unknown dynamical mechanism to drive the number to $w = -2/3$,
or an uncomfortable fine tuning. We can see that, for a universe
with a decreasing energy density, the denominator in the second term
continues to decrease with time and drives the second term away
from the phenomenologically acceptable value $-1 < w < 1$,
unless $w(t_i)$ in the numerator is fine tuned to be arbitrarily
close to $-2/3$.

We wish to comment that the difficulty we find in attempting to
stabilize the radius of the extra dimensions to a fixed value is similar
to the problem of trying to have a static Einstein universe.
Both require fine tuning. While a static universe can be achieved
with a proper parameter choice, it is unstable to perturbations
about these values; similarly, a stabilized radius of the extra dimension
is unstable to slight deviation of $w(t_i)$ from the precise value
of 2/3.  Although we have only been able to prove the fine tuning
required for radius stabilization for specific metrics in the
Ho\v{r}ava Witten scenario, we suspect that such a result may generalize
to any attempt to stabilize the radius in the context of an extra
dimension in which gravitons propagate\footnote{Remarks on the difficulty
of stabilizing the radius in extra dimensions can also be 
found in \cite{steinhardt}, which appeared shortly after this work}.

A way to avoid these conclusions and still to satisfy \eqr{eq:naive}
is if $\rho(t)$ on the observable 3-brane is dominated by a constant
vacuum energy (an effective cosmological constant) for all time.  Then
$\rho(t) = \rho(t_i)$, $w(t) = w(t_i) = -1$, and \eqr{eq:naive} is 
trivially satisfied.  Such an energy density on the brane does not
automatically lead to exponential expansion of the brane if there
is a nonlocal contribution canceling the effect of the
3-brane cosmological constant, and hence this case can possibly be
made to match today's universe.  However, generically \eqr{eq:naive}
disagrees with the evolution of our observable universe.

The condition that the metric on the orbifold be absolutely
static is clearly too stringent. The radius of the orbifold
must be allowed to change.  Then the only experimental
constraint on the variation of the orbifold length comes from the bound on
the variation of the Newton's constant (see for example
\cite{Degl'Innocenti:1996nt} and references therein).  The most
conservative bound is approximately
\begin{equation}
\frac{d G_N}{dt}/G_N  < 6 \times 10^{-11} \mbox{yr}^{-1}  \, ,
\label{eq:newtonbound}
\end{equation}
which comes from the spin down rate of pulsars.  Although this bound is
only measured  on astrophysical time scales, we will assume it
to be valid for cosmological time scales.  We remind the reader
that $G_N$ is inversely proportional to the distance between
3-branes (\eqr{eq:one}), which is obtained via \eqr{eq:orblength}
as an integral over the entire bulk.  Hence the best way to avoid
the constraints of \eqr{eq:staticbulk} and \eqr{eq:naive}, while still
satisfying the observations on $dG_N/dt$, is to have a large
time variation of the orbifold metric near the brane and vanishing
time variation elsewhere throughout the bulk.  We will construct such
an example shortly.

The most general constraint from \eqr{eq:israelnu} arises as a result
of $w$ being of order unity.  This implies that, at $u=0$,
\begin{equation}
| e^{-\beta} \beta' l | \sim \epsilon \equiv |t^0_0 l| \sim 10^{-114} 
\label{eq:flatbeta}
\end{equation}
where $l \sim 10^{-15} \mbox{GeV}^{-1}$ is the physical length scale
of the orbifold appropriate for the Ho\v{r}ava-Witten model considered in
\cite{ovrut}. The numerical estimate in \eqr{eq:flatbeta} is made by
taking the energy
density to be $\rho \sim 10^{-29}$ gm cm$^{-3}$ as can be
roughly inferred from rotation curves of galaxies (we motivate this
number with rotation curves rather than with the typical critical
density of standard cosmology because the latter may change in these
models).  This condition says that, very near the brane, the orbifold
scale factor remains approximately independent of the transverse distance
off the brane.  Below we will allow the physical length scale of the
orbifold $l$ to be as large as a $mm$ and find that our conclusions
are relatively unchanged.

Now we will construct an explicit example which demonstrates
that the time variation bound and \eqr{eq:flatbeta} can be
simultaneously satisfied by allowing the orbifold length to have a
time variation.  However, as we will see, the model is still
unnatural.  Consider the following ansatz:
\begin{equation}
\label{eq:ansatz}
\beta \equiv \frac{2 \tau h}{\lambda} e^{- \lambda l/2} \mbox{cosh}
\left[ \lambda ( u -  l/2) \right]
\end{equation}
where $\lambda l \gg 1$. 
Note that this ansatz fits the optimal situation described below 
\eqr{eq:newtonbound} in that most of the time variation is near
the boundary: $dG_N/dt$ is determined by $d\beta/dt $ with $\beta$ (and
its time derivative) biggest at the boundaries.  
We define the orbifold proper length to be
\begin{equation}
\label{eq:leng}
\int_0^l e^{\beta(\tau(t), u)} du
\end{equation}
where $t$ is the comoving time on the brane.  
Then, from equations \eqr{eq:one}, \eqr{eq:newtonbound},
\eqr{eq:ansatz}, and \eqr{eq:leng}, the time variation bound on the
Newton's constant translates into
\begin{equation}
\frac{2 | h|}{l \lambda^2} < c_1
\label{eq:newtontime}
\end{equation}
where $c_1 = 6 \times 10^{-11} \mbox{yr}^{-1}$.  Dropping factors of
order unity, combining \eqr{eq:newtontime} and \eqr{eq:flatbeta}, we
find the constraint on $h$ to be
\begin{equation}
\frac{2}{c_1 \epsilon^2 t^2 l} < |h|
\end{equation}
where $\epsilon =10^{-114}$ for $l \sim 10^{-15} \mbox{GeV}^{-1}$.
For example, if one takes $t = 10^{10}$yr, then if $|h| > 10^{233}$
GeV/yr, the time variation constraint on the Newton's constant is
satisfied.  Note that if one took $l$ as large as 1 mm, the parameter
will still be unnatural; since $\epsilon \propto l$,
the bound on $h \propto 1/l^3$ and we would require $|h|
>10^{149}$ GeV/yr. Of
course, such a large dimensionful constant most likely is unphysical.
Incidentally, for $l \sim 10^{-15} \mbox{GeV}^{-1}$, $\lambda >
10^{127}$ GeV which is much greater than $1/l$ as was assumed below
\eqr{eq:ansatz}.  Note also that although \eqr{eq:newtontime} seems to
imply an upper bound on the value of $|h|$, instead a lower bound
arises because $\lambda$ in \eqr{eq:flatbeta} has to adjust as $|h|$
is varied to keep the equation of state reasonable.  This requirement
of such an unphysically large number for $h$ is a
result of $t^0_0 l$ being very small.  This in
turn is just a restatement of the cosmological constant problem in the
sense that there is a severe mismatch between the natural scale
$1/(\kappa_5^{2} l)$ and the energy density on the brane.
One way of accomodating the severe mismatch is to have the long age of
the universe (corresponding to the small energy density of the
universe) built into the solution itself.  Indeed, Eqs.\
(\ref{eq:examplesol1}) and (\ref{eq:examplesol2}) with $q=1$ give such
an example, where $1/\tau \sim {\cal O}(1/10^{10} \mbox{yr})$ supplies
the unnaturally small (compared to $1/(\kappa_5^2 l )^{1/4}$) energy scale.

{\it A cosmological example of our 3-brane for $\nu=\beta$}

Now we turn to evolution of the 3-brane when 
$\nu = \beta$.
We consider the other Israel equation \eqr{eq:israelalpha} which
specifies $\rho$ once $\alpha$ is given.  Suppose we have the
brane expanding as a power law in comoving time.  Thus, we write
\begin{equation}
\label{eq:alpha}
\alpha= \ln f(u) + m \ln(t(\tau, u)/t_i)
\end{equation}
where $m$ is the exponent of the power law and $t_i$ is a
constant.  Using the ansatz for $\beta$ in \eqr{eq:ansatz}, we then find 
\begin{equation}
\rho = A(u) \frac{1}{t} \ln(1+ t \theta) + B(u) \frac{1}{1+ t \theta }
\end{equation}
where $A(u)$ and $B(u)$ are some functions of $u$ and
\begin{equation}
\theta \equiv \beta/\tau.
\end{equation}
Note that, as $t \rightarrow \infty$, $\rho \rightarrow (A + B/\theta)
/t \propto H$.  This is in contrast with the FRW cosmology where $\rho
\propto 1/t^2 \propto H^2$
independently of the equation of state, as long as the scale factor is
a power law with respect to time.  Hence, we have a novel feature of
the energy density scaling like the Hubble speed instead of its
squared.  As discussed in the previous section, such a relationship
between the energy density and the time causes many cosmological
problems.  Furthermore, if one writes down the equation of state 
corresponding to this cosmology, one will see that $w$ approaches a 
constant value that is determined, e.g., by the energy
density at some initial time.

Note that this
behavior relies strongly on the assumption $\nu=\beta$, which cannot
always be made to be true if the orbifold fixed planes have time
independent coordinate positions.  Indeed, without the choice
$\nu=\beta$, the physical time derivative taken to obtain the Hubble
speed has no general relationship with the physical spatial (orbifold
directional) derivative taken to obtain the energy density $\rho$.
Note that the bulk equations are largely unconstrained because of the
freedom we have in choosing $\alpha(\tau, u)$ for $u$ away
from the boundaries.  Even with
the simple
power law expansion of the scale factor discussed in this section, $f(u)$
in \eqr{eq:alpha} is unconstrained except at $u=0$.  

\section{Causality and the Horizon Problem}

Interesting new avenues for solutions to the horizon problem
arise in the context of extra dimensions.  Here we confine
ourselves to a few remarks and reserve a more complete discussion
to a work in progress.  Previous work by Freese and Levin \cite{levin1}
addressed the horizon problem in some generality, of course in a 
four-dimensional context.  We discuss here one aspect of the generalization
of that work to higher dimensions.  The major point we wish
to make is that the application of standard inflation requirements,
such as the requirement of 60 e-foldings, may be modified
in the context of extra dimensions as some of the assumptions
that went into obtaining this requirement are relaxed.

Note that the discussion in this section is true for any model in
which the length of the extra dimension(s) determines the
four-dimensional gravitational coupling.  For example, it applies
to periodic boundary conditions considered by many authors
including \cite{dimopoulos}.  

An estimate of the size of the observable universe today
is given by the distance light could travel between photon
decoupling and now, $d_{obs} \sim a(t_o) \int_{t_{dec}}^{t_o}
dt/a(t) = O(1) \times (t_o - t_{dec})$ for $a \propto t^p$
and $p=O(1)$ between $t_{dec}$ and $t_o$.  Here $a$ is the
scale factor. We can compare the comoving size of the observable
universe to the comoving size of a causally connected region at
some earlier time $t_c$: $d_{hor}(t_c)/a(t_c)
= \int_0^{t_c} {dt\over a(t)}.$  The observable universe today
fits inside a causally connected region at $t_c$ if
\begin{equation}
{d_{hor}(t_c) \over a(t_c)} \geq {d_{obs} \over a(t_o)} \, .
\label{eq:causality}
\end{equation}
Here, subscript $o$ refers to today and subscript $c$ to some early
time.  If condition \eqr{eq:causality} is met, then the horizon size
at $t_c$ (before nucleosynthesis) is large enough to allow for a
causal explanation of the smoothness of the universe today.  Note that
more creative explanations of large scale smoothness may not involve
comparing these two patches.
For example, in the context of the brane scenarios, one might imagine
that two regions of our observable universe which seem to be causally
disconnected might in fact have talked to each other because of a geodesic
between them that went off our brane, into the bulk, and then back
onto our brane at some distant point. Such alternatives are certainly
worth investigating further. In the remainder of our discussion
here we restrict
ourselves to the case where \eqr{eq:causality} is relevant; this is
certainly the case for all the brane and boundary inflation models
considered to date.

For power law expansion of the scale factor both before $t_c$ and
after $t_{dec}$ (which may or may not be the case), we can take
$t \sim H^{-1}$ during these periods. The causality condition
\eqr{eq:causality} then becomes
\begin{equation}
{1 \over a_c H_c} \geq {1 \over H_o a_o} \, .
\label{eq:causality2}
\end{equation}
We take the Hubble constant today to be 
\begin{equation}
H_o = \alpha_o^{1/2}T_o^2/m_{pl}(t_o),
\label{eq:htoday}
\end{equation}
where $\alpha_o = (8 \pi/3)(\pi^2/30)g_*(t_o)\eta_o$,
$g_*$ is the number of relativistic degrees of freedom and
$\eta(t_o) \sim 10^4-10^5$ is the ratio today of the energy
density in matter to that in radiation.

Following Freese and Levin \cite{levin1}, one can further simplify
this condition by making a series of simplifying assumptions.
First, we assume for now that the entropy on the 3-brane both at time
$t_c$ and today obeys $S \propto (aT)^3$
such that we can write $a$ in \eqr{eq:causality} in terms of the
entropy.  A work in progress is reexamining this condition.

A further assumption is always made in obtaining requirements for
inflation to succeed.  It is that the Hubble speed at the early
time $t_c$ is given by
\begin{equation}
H_c \sim T_c^2/m_{pl}(t_c),
\label{eq:inithubble}
\end{equation}
where $T_c$ is some mass scale, possibly the temperature, at time
$t_c$.  As can be seen by the earlier results in this current paper,
this assumption may not always hold.  Thus one should in principle
always go back to the original equation \eqr{eq:causality} to check
that any particular inflation proposal really does solve the horizon
problem.  
However, for the models that have been heretofore proposed, the usual
requirements used are at least qualitatively similar to the actual
requirements implied by above equations.  

Even if above assumptions hold, yet another fact must be considered:
the changing four dimensional Planck mass.
Using Eqs.\ (\ref{eq:causality2}), (\ref{eq:htoday}),
(\ref{eq:inithubble}), and the entropy relation, 
the causality condition becomes
\begin{equation}
\label{eq:changegrav}
{m_{pl}(t_c) \over m_{pl}(t_o)} \bigl({S_o \over S_c} \bigr)^{1/3}
\geq 10^{-2} {T_c \over T_o} \, .
\end{equation}
One can now invoke the idea of changing the four dimensional Planck mass
in lieu
of changing the entropy in order to solve the horizon
problem, as has been suggested by Levin and Freese \cite{levin}
and as is studied in the context of string theory by
Veneziano \cite{veneziano} and collaborators.  Here, in the context of extra
dimensions, if the compactified dimension that determines
the strength of gravity shrinks with time, one may be 
able to use the changing four dimensional Planck mass to explain causality
as in \eqr{eq:changegrav} (see also Riotto \cite{riotto}).

In the context of inflation, one must also be careful about the
changing four dimensional Planck mass due to the changing size of the
extra dimension
in obtaining the appropriate causality condition.  Suppose inflation
ends at time $t_e$.  Assuming that entropy increases by the end of
inflation while it remains constant after inflation, we can then
rewrite the condition \eqr{eq:changegrav} as
\begin{equation}
{a_e \over a_c} \geq 10^{-2} {T_c^2 \over T_e T_o} 
\bigl( {m_{pl}(t_o) \over m_{pl}(t_c)} \bigr) \, .
\end{equation}
It is this last factor of $m_{pl}(t_o)/m_{pl}(t_c)$
that must not be forgotten, if the value
of the four dimensional Planck mass changes between the time of inflation
and now.
One way to think about this is that in taking our observable
universe today and rescaling it to some early time (to see
if it fits inside a causal radius), the size of the observable
patch may be bigger (or smaller) than one would naively think because of the
difference in the Planck masses.  
For example, in the inflation model of \cite{dimopoulos}
inflation takes place at an early stage while the radius of compactification
is very small and the Planck mass much smaller than today.
If one takes $m_{pl}(t_c) =$TeV, for example, then one
needs the scale factor to grow
by an additional factor of $10^{19} {\rm GeV}/{\rm TeV} =
10^{16}$ compared to the standard inflationary
result with constant Planck mass.
Instead of the usual 60 e-foldings one needs 100. 
This is not a significant difference, since it is usually just as easy
to get 100 e-foldings as 60. Still, we caution that there
are many assumptions that go into the calculation of requirements
for inflationary models, and these should in principle be checked.

\section{Conclusions}
\label{sec:conc}
In this work, we have explored the general cosmology of a flat, homogeneous
and isotropic 3-brane located at the $Z_2$ symmetry fixed plane of a
$Z_2$ symmetric five dimensional spacetime as in the Ho\v{r}ava-Witten
model compactified on a Calabi-Yau manifold.  We have found an important
manifestation of the moduli problem: the usual relationship between
the FRW scale factor and the energy density on our 3-brane is modified
from $\rho \sim H^2$ to almost any relationship between $\rho$ and
$H$, such as $\rho \sim H^q$ with $q \neq 2$.  Hence, in these models,
one does not generically recover the standard FRW evolution for our
observable universe.  The time evolution of our observable universe
can be controlled by the moduli from the extra dimensions and the
attending boundary conditions of the spacetime, rather than by the
energy density on our brane.  These modifications persist even in the
limit that the bulk stress energy vanishes, because of the
modifications to Einstein's equations in the bulk and the nontrivial
Israel conditions arising from the $Z_2$ symmetry.  We have shown that
this typically leads to difficulties with primordial nucleosynthesis,
structure formation, and bounds on the age of the universe.

We have also found that, in a large class of cases,
it is impossible
to stabilize the radius of the extra dimension (i.e., ``pin" the
4-dimensional Planck mass at a fixed value) without fine tuning
the equation of state on our brane.
The difficulty we find in attempting to
stabilize the radius of the extra dimensions to a fixed value is similar
to the problem of trying to have a static Einstein universe.
We find that the radius of the extra dimension can be frozen
at a fixed value in a universe (on the 3-brane) with decreasing energy
density  only if the equation of state satisfies $w = P/\rho = -2/3$;
however, the radius is unstable to slight deviations from this value
such that $w$ must be fine-tuned to extreme precision to exactly
this number. Although this particular value has recently been discussed,
e.g. by Wang {\it et al} \cite{wang} as tentatively favored by overlap of
a number of astrophysical observations, the lack of a dynamical mechanism
to drive $w$ to this value renders this exact number extremely fine tuned.
Even a slowly varying Planck mass, in agreement with observations,
requires a tuning.
This tuning is a result of the energy density of the universe
today being severely mismatched with the energy density set by the scale
of the fundamental Planck constant.
Although we have only been able to prove the fine tuning
required for radius stabilization for specific metrics in the
Ho\v{r}ava-Witten scenario
(those with conformally flat metric in the $\tau-u$ subspace), 
we suspect that such a result may generalize
to any attempt to stabilize the radius in the context of an extra
dimension in which gravitons propagate. 

We repeat, however, that this constraint involving the equation of
state can be evaded if one assumes that the energy density confined to the
3-brane, $\rho$, is dominated by the cosmological constant.  In that
case, the equation of state as well as the energy density can remain
constant in time.  A cosmological constant
which dominates the energy density $\rho$ confined to the brane does not
necessarily lead to an inflationary universe; in principle there can
be other (nonlocal) contributions from off the brane canceling
the effects of the cosmological constant, such that the evolution
of our universe today can proceed as usual.
The ordinary matter in this case
would contribute only negligibly to $\rho$. 

We can apply the results of our paper to the model of Randall and
Sundrum \cite{sundram}, which falls into the class of models
we have considered here. First, our result that FRW cosmology does not
generically result on our brane would apply in their model also.
Second, they presumably avoid the problems of stabilizing the orbifold
radius by having a cosmological constant dominate the energy density
of our brane, as described in the previous paragraph.  

Finally, we made some remarks about causality and the horizon
problem in theories with extra dimensions.  We discussed possible
modifications to the standard inflationary picture.  For example,
the time variation of the gravitational
coupling due the changing size of the compactified dimension
can be used to solve the horizon problem or at least modify
the number of e-foldings needed to solve the horizon problem within
the context of inflation.  In addition, the modifications indicated here 
to the standard FRW relation 
$H^2 \propto \rho$ would change the requirements of 
inflationary models.

\acknowledgements{We thank Pierre Binetruy, Michael Bershadsky, Marty
Einhorn, Dan Kabat, Dan Waldram, Josh Frieman, Rocky Kolb, Lisa
Everett, Paolo Gondolo, and Gus Evrard 
for useful discussions. We acknowledge support
from the Department of Energy at the University of Michigan Physics
Dept. K. Freese thanks the theory group at CERN in Geneva and the Max
Planck Institut fuer Physik in Munich, where part of this work was
completed, for hospitality during her stay.}



\begin{thebibliography}{99}
\frenchspacing
\def\prpts#1#2#3{Phys. Reports {\bf #1}, #2 (#3)}
\def\prl#1#2#3{Phys. Rev. Lett. {\bf #1}, #2 (#3)}
\def\prd#1#2#3{Phys. Rev. D {\bf #1}, #2 (#3)}
\def\prc#1#2#3{Phys. Rev. C {\bf #1}, #2 (#3)}
\def\plb#1#2#3{Phys. Lett. {\bf #1B}, #2 (#3)}
\def\npb#1#2#3{Nucl. Phys. {\bf B#1}, #2 (#3)}
\def\apj#1#2#3{Astrophys. J. {\bf #1}, #2 (#3)}
\def\apjl#1#2#3{Astrophys. J. Lett. {\bf #1}, #2 (#3)}
\bibitem{horavawitten} P.~Ho\v{r}ava and E.~Witten,
\npb{460}{1996}{506}; P.~Ho\v{r}ava and E.~Witten,
\npb{475}{1996}{94}; P.~Ho\v{r}ava,
\prd{54}{1996}{7561} E.~Witten,
\npb{471}{1996}{135}

\bibitem{cosmobefore}
T.~Nihei,
``Inflation in the five-dimensional universe with an orbifold extra
                  dimension,''
hep-ph/9905487;
K.~Benakli,
``Cosmological solution in M theory on S-1 / Z(2),''
Int. J. Mod. Phys. {\bf D8}, 153 (1999)
hep-th/9804096;
N.~Arkani-Hamed, S.~Dimopoulos and G.~Dvali,
``Phenomenology, astrophysics and cosmology of theories with submillimeter
                  dimensions and TeV scale quantum gravity,''
Phys. Rev. {\bf D59}, 086004 (1999)
hep-ph/9807344;
T.~Banks, M.~Dine and A.~Nelson,
``Constraints on theories with large extra dimensions,''
hep-th/9903019;
G.~Dvali and S.H.~Tye,
``Brane inflation,''
Phys. Lett. {\bf B450}, 72 (1999)
hep-ph/9812483;
N.~Kaloper and A.~Linde,
``Inflation and large internal dimensions,''
Phys. Rev. {\bf D59}, 101303 (1999)
hep-th/9811141;
N. ~Kaloper, ``Bent domain walls as braneworlds,"
hep-ph/9905210;
D.H.~Lyth,
``Inflation with TeV scale gravity needs supersymmetry,''
Phys. Lett. {\bf B448}, 191 (1999)
hep-ph/9810320;
E. ~Halyo, ``D-Term Inflation at the TeV scale and 
large internal dimensions," hep-ph/9905244;
James ~Cline, ``Inflation from extra dimensions," hep-ph/9904495.

\bibitem{chamblin}  H. A. Chamblin and H. S. Reall, hep-th/9903225.

\bibitem{ovrut} Andr\'e Lukas, Burt A. Ovrut, and Daniel Waldram,
hep-th/9902071.

\bibitem{binetruy} P.~Binetruy, C.~Deffayet and D.~Langlois,
``Nonconventional cosmology from a brane universe,''
hep-th/9905012.

\bibitem{csaki} C. Csaki, M. Graesser, C. Kolda, and J. Terning,
``Cosmology of one extra dimension with localized gravity,"
hep-ph/9906513.

\bibitem{cline2} J.M. Cline, C. Grojean, and G. Servant,
Phys. Rev. Lett. (to be published), hep-ph/9906523.

\bibitem{riotto}
A.~Riotto,
``D-branes, string cosmology and large extra dimensions,''
hep-ph/9904485.

\bibitem{chaboyer}
B.~Chaboyer, P.~Demarque, P.J.~Kernan and L.M.~Krauss,
``The Age of globular clusters in light of Hipparcos: Resolving the
age problem?'', astro-ph/9706128.

\bibitem{wang}
L. ~Wang, R.R. ~Caldwell, J.P. ~Ostriker, and P.J. ~Steinhardt,
``Cosmic Concordance and Quintessence,"
astro-ph/9901388.

\bibitem{steinhardt}
P.J. ~Steinhardt, ``General considerations of the
cosmological constant and the stabilization of moduli
in the brane-world picture," hep-th/9907080.

\bibitem{Degl'Innocenti:1996nt}
S.~Degl'Innocenti, G.~Fiorentini, G.G.~Raffelt, B.~Ricci and A.~Weiss,
``Time variation of Newton's constant and the age of globular
clusters,''
Astron. Astrophys. {\bf 312}, 345 (1996)
astro-ph/9509090.

\bibitem{levin1}
K.~Freese and J.~Levin,
``Beyond Potential Dominated Inflation,"
Proceedings of Conference on Unified Symmetry in Coral Gables, FL, 1994.

\bibitem{dimopoulos}
N.~Arkani-Hamed, S.~Dimopoulos, N.~Kaloper and J.~March-Russell,
``Rapid asymmetric inflation and early cosmology in theories with
                  submillimeter dimensions,''
hep-ph/9903224.

\bibitem{levin}
J.J.~Levin and K.~Freese,
``A Possible solution to the horizon problem: The MAD era for massless
scalar
                  theories of gravity,''
Phys. Rev. {\bf D47}, 4282 (1993)
astro-ph/9211011.

\bibitem{veneziano} G.~Veneziano,
``Scale factor duality for classical and quantum strings,''
Phys. Lett. {\bf B265}, 287 (1991).

\bibitem{sundram}
L. ~Randall and R. ~Sundrum,
``A Large Mass Hierarchy from a small extra dimension,"
hep-ph/9905221.

\end{thebibliography}
\end{document}